\documentclass[12pt]{article}
\usepackage{a4,amsmath}
\usepackage{epsfig}
\begin{document}
\begin{Large}
\noindent
{\bf{On relativistic symmetry of Finsler spaces\\[2mm]
  with mutually opposite preferred directions}}
\end{Large}
\vskip 6mm
\begin{large}
\noindent
{\bf{G.\,Yu. Bogoslovsky\\}}
\end{large}
\noindent
Skobeltsyn Institute of Nuclear Physics\\
Lomonosov Moscow State University,  Russia\\
E-mail\,: bogoslov@theory.sinp.msu.ru\\
\vskip 5mm
\begin{small}
\noindent
It is shown that in Minkowski space there exist transformations of the coordinates of events alternative to the 3-parameter Lorentz boosts. However, in contrast to the boosts, they constitute a 3-parameter noncompact group which, in turn, is a subgroup of the homogeneous 6-parameter Lorentz group. Moreover, in the same space, there exists another 3-parameter noncompact group isomorphic to above-mentioned one. As we shall see, these two 3-parameter noncompact groups are rudiments of the 3-parameter groups of relativistic sym\-metry of the axially symmetric Finslerian spaces with the preferred directions $\boldsymbol\nu$  and $-\boldsymbol\nu$, respectively. Finally, it will be also demonstrated that inversion of the preferred direction $\boldsymbol\nu$ in the axially symmetric Finslerian space-time does not change the Lobachev\-ski geometry of 3-velocity space. However, this leads to an inversion of the corresponding family of horospheres of the space.\\
\end{small}
\vskip 5mm
\noindent
{\large{\bf 1\,. Introduction}}

\bigskip

\noindent
As it is known, space-time is Riemannian within the framework of GR, and the distribution
and motion of matter only determines the local curvature of space-time without
affecting the geometry of the tangent spaces. In other words, regardless of the properties
of the material medium which fills the Riemannian space-time, any flat tangent space-time
remains the space of events of SR, i.e. the Minkowski space with its Lorentz symmetry,
which is usually identified with the relativistic symmetry.

However, in recent literature there is an increasing interest in the problem of violation
of Lorentz symmetry (\,see [1] and the references cited therein\,). Particularly, the string-motivated approach to this problem is widely discussed.

The point is that even if the original unified theory of interactions possesses Lorentz
symmetry up to the most fundamental level, this symmetry can be spontaneously broken
due to the emergence of the condensate of vector or tensor field. The appearance of
such a condensate, or of a constant classical field on the background of Minkowski space,
implies that it can affect the dynamics of the fundamental fields and thereby modify the
Standard Model of strong, weak and electromagnetic interactions. Since the constant
classical field is transformed by the passive Lorentz transformations as a Lorentz vector
or tensor, its influence on the dynamics of fundamental fields of the Standard Model is
described by the introduction of the additional terms representing all possible Lorentz-covariant
convolutions of the condensate with the Standard fundamental fields into the
Standard Lagrangian. The phenomenological theory, based on such a Lorentz-covariant modification of the Standard model is called the Standard Model Extension (\,SME\,) [2].

By design, the phenomenological SME theory is not Lorentz-invariant, since its Lagrangian
is not invariant under active Lorentz transformations of the fundamental fields
against the background of fixed condensate. In addition, in the context of SME, a violation
of Lorentz symmetry also involves the violation of relativistic symmetry, since
the presence of non-invariant condensate breaks the physical equivalence of the different
inertial reference systems.

It should be noted that in the low-energy limit of gravitation theories with broken
Lorentz and relativistic symmetries, there appears an unlimited number of possibilities to
build a variety of effective field theories, each of which being potentially able to explain at least
some of the recently discovered astrophysical phenomena. At the same time, the very existence
of the Finsler geometric models of space-time [3],\,[4] within which a violation
of Lorentz symmetry occurs without the violation of relativistic symmetry strongly
constrains the possible effective field theories with broken Lorentz symmetry: in order
to be viable, such theories, in spite of the presence of Lorentz violation, should have the
property of relativistic invariance.

Note also that, as shown in [4], the Ridge/CMS-effect revealed at the Large Hadronic Collider,
directly demonstrates that in the early Universe there
spontaneously emerged the axially symmetric local anisotropy of space-time with a group
DISIM${_{b}}$(2) as an inhomogeneous group of local relativistic symmetry and the corresponding
Finsler metric
\begin{equation}\label{1}
ds^2=\left[\frac{(dx_0-\boldsymbol\nu d\boldsymbol x)^2}{dx_0^2-d\boldsymbol x^{\,2}}\right]^r
(dx_0^2-d\boldsymbol x^{\,2})\,.
\end{equation}
This metric, proposed for the first time in [5], depends on two constant parameters $r$ and $\boldsymbol\nu$, and generalizes the Minkowski metric. Here $r$ determines the magnitude of spatial anisotropy, characterizing, thus, the degree of deviation of (1) from the isotropic Minkowski metric. Instead of the 3-parametric group of rotations of Minkowski space, Finsler spaces (1) allow only one 1-parameter group of rotations around the unit vector $\boldsymbol\nu$, which represents a physically preferred direction in 3D space.

One of the most important distinguishing features of Finsler spaces (1) consists in their noninvariance under the discrete improper transformations: $x_0\rightarrow -x_0$ or $\boldsymbol x\rightarrow -\boldsymbol x.$ This suggests that the emergence of the axially symmetric local anisotropy of space-time in the early Universe should be accompanied by violation of the CPT invariance (\,in this connection see [6]\,). In order to scrutinize such a problem, we shall take here the first steps towards the objective, namely consider inhomogeneous groups of local relativistic symmetry of Finsler spaces (1) and of
\begin{equation}\label{2}
ds^2=\left[\frac{(dx_0+\boldsymbol\nu d\boldsymbol x)^2}{dx_0^2-d\boldsymbol x^{\,2}}\right]^r
(dx_0^2-d\boldsymbol x^{\,2})\,.
\end{equation}
Obviously, the Finsler space (2) is obtained from (1) by replacing \,$x_0\,\rightarrow\, -x_0$\, or\, $\boldsymbol x\,\rightarrow\, -\boldsymbol x\,.$
\vskip 5mm
\noindent
{\large{\bf 2\,. Axially symmetric Finsler spaces and their isometry groups \phantom{asai}as inhomogeneous groups of local relativistic symmetry}}

\bigskip

\noindent
For a start let us consider the flat Finsler space-time (1). As to the isometry group of (1) and to its Lie algebra, for the first time they were found (in an explicit form) in [7],\,[8],\,[9]. The respective group turned out to be 8-parametric: four parameters correspond to space-time translations, one parameter, to rotations about  preferred direction $\boldsymbol\nu$, and three parameters, to the generalized Lorentz boosts. One should notice that at present, after the works [10],\,[11], this 8-parameter group is increasingly referred to as DISIM${_{b}}$(2), where $b$  is the new designation of the above-mentioned parameter $r$ (\,for more details concerning what has been said, see, in particular, [12],\,[13]\,). As to
the abbreviation DISIM${_{b}}$(2), this stands for Deformed Inhomogeneous SIMilitude group that includes a 2-parameter Abelian homogeneous noncompact subgroup. Nevertheless, hereafter we shall hold on to our original designations.

Now let us consider infinitesimal transformations of the 8-parameter isometry group of the axially symmetric Finsler space-time (1). Originally (\,see [7]\,), the corresponding transformations of its  3-parameter homoge\-neous noncompact subgroup, i.e. infinitesimal transformations of relativi\-stic symmetry of space-time (1), were obtained in the form
\begin{eqnarray}\label{3}
dx_0\!\!\!\!\!&=&\!\!\!\!(\,-\, r\,(\boldsymbol{\nu}\boldsymbol{n})x_0 - \boldsymbol{n} \boldsymbol{x}\,)\, d \alpha\,, \nonumber\\
d\boldsymbol{x}\!\!\!\!&=&\!\!\!\!(\,-\, r\,(\boldsymbol{\nu}\boldsymbol{n})\boldsymbol{x} - \boldsymbol{n} x_0 - [\boldsymbol{x}[\boldsymbol{\nu}\boldsymbol{n}]]\,)\,d\alpha\,,
\end{eqnarray}
where the unit vector $\boldsymbol n$ and  $\alpha $ are the group parameters.
As to the infinitesimal transformations of the above-mentioned 1-parameter group
of rotations and of the 4-parameter group of space-time translations, they have the form
\begin{equation}\label{4}
d\boldsymbol{x}=\, [\boldsymbol x\boldsymbol\nu ]\,d\omega\,; \quad\, dx^i=da^i\,, \quad\, i=0,\,1,\,2,\,3\,.
\end{equation}
Using all these infinitesimal transformations with the condition that the third space axis is directed along $\boldsymbol\nu$ and three successive directions (along the first-, the second- and the third axis) are chosen for $\boldsymbol n$, we arrive at the simplest representation of generators of the 8-parameter isometry group of the Finsler space-time (1). As a result,
\begin{equation}\label{5}
\left.
\begin{array}{rcl}
X_1\!\!&\!\!=\!\!&\!\!-(x^1p_0+x^0p_1)-(x^1p_3-x^3p_1)\,,\\
X_2\!\!&\!\!=\!\!&\!\!-(x^2p_0+x^0p_2)+(x^3p_2-x^2p_3)\,,\\
X_3\!\!&\!\!=\!\!&\!\!-rx^ip_i-(x^3p_0+x^0p_3)\,,\\
R_3\!\!&\!\!=\!\!&\!\!x^2p_1-x^1p_2\,; \qquad \qquad \qquad \qquad \qquad p_i=\partial /\partial
x^i\,.\\
\end{array}
\right.
\end{equation}
These generators satisfy the following commutation relations\,:
\begin{equation}\label{6}
\left.
\begin{array}{llll}
  [X_1X_2]=0\,, & [R_3X_3]=0\,, && \\
  \left[X_3X_1\right]=X_1 \,, & [R_3X_1]=X_2 \,, && \\
  \left[X_3X_2\right]=X_2\,, & [R_3X_2]=-X_1\,; & &  \\
\end{array}
\right.
\end{equation}

\vspace*{-5mm}
\begin{equation}\label{7}
\left.
\begin{array}{llll}
  \left[p_i p_j\right]=0\,; &&& \\
  \left[X_1p_0\right]=p_1\,,& [X_2p_0]=p_2\,, & [X_3p_0]=rp_0+p_3\,, &
  [R_3p_0]=0\,, \\
  \left[X_1p_1\right]=p_0+p_3\,,& [X_2p_1]=0\,, & [X_3p_1]=rp_1\,, &
  [R_3p_1]=p_2\,, \\
  \left[X_1p_2\right]=0\,, & [X_2p_2]=p_0+p_3\,, & [X_3p_2]=rp_2\,, &
  [R_3p_2]=-p_1\,, \\
  \left[X_1p_3\right]=-p_1\,, & [X_2p_3]=-p_2\,, & [X_3p_3]=rp_3+p_0\,, &
  [R_3p_3]=0\,. \\
\end{array}
\right.
\end{equation}
The operators $X_1,\,X_2,\,X_3$ and their Lie algebra correspond to the special case where the third spatial axis is directed along $\boldsymbol\nu$.
In the general case where spatial axes are oriented arbitrarily with respect to the preferred direction, the corresponding operators generate the following finite homo\-geneous transformations (\,the generalized Lorentz boosts making up the 3-parameter group of relativistic symmetry of the flat axially symmetric Finslerian event space (1)\,)\,:
\begin{equation}\label{8}
x'^i=D(\boldsymbol v,\boldsymbol\nu )R^i_j(\boldsymbol
v,\boldsymbol\nu )L^j_k(\boldsymbol v) x^k\,,
\end{equation}
where $\boldsymbol v$ denotes the velocities of moving (primed)
inertial reference frames, the matrices $L^j_k(\boldsymbol v)$
represent the ordinary Lorentz boosts, the matrices
$R^i_j(\boldsymbol v,\boldsymbol\nu )$  represent additional
rotations of the spatial axes of the moving frames around the
vectors $[\boldsymbol v\boldsymbol\nu ]$ through the angles
\begin{equation}\label{9}
\varphi=\arccos\left\{ 1-\frac{(1-\sqrt{1-\boldsymbol v^{2}/c^2})
[\boldsymbol v\boldsymbol\nu ]^2}{ (1-\boldsymbol v\boldsymbol\nu
/c)\boldsymbol v^{2}}\right\}
\end{equation}
of the relativistic aberration of $\boldsymbol\nu,$ and the diagonal matrices
\begin{equation}\label{10}
D(\boldsymbol v,\boldsymbol\nu )=\left(\frac{1-\boldsymbol
v\boldsymbol\nu /c} {\sqrt{1-\boldsymbol v^{2}/c^2}} \right)^rI
\end{equation}
stand for the additional dilatational transformations of the event coordi\-nates.

Note that the structure of the generalized Lorentz boosts (8) ensures the fact that, in spite of new (\,Finsler\,) geometry of the flat event space (1), the 3-velocity space remains to be a Lobachevski space (\,see, for instance [14]\,) with metric
\begin{equation}\label{11}
dl^2_{v}=\frac{(d\boldsymbol v)^2-[{\boldsymbol v}d{\boldsymbol v}]^2}{(1-{\boldsymbol v}^2)^2}\,.
\end{equation}
Thus, the transition from the Minkowski event space to the flat axially symmetric Finsler event space (1) leaves the relativistic 3-velocity space unchanged. Therefore it is clear that the 3-parameter group of the generalized Lorentz boosts (8) induces an isomorphic 3-parameter group of the correspon\-ding motions of the Lobachevski space. In particular, the Abelian (\,see (6)\,) 2-parameter subgroup with
the generators $X_1\,, X_2$  (\,see (5)\,) induces a 2-parameter subgroup of such motions of the Lobachevski space which leave invariant a family of the horospheres $(1\!- \!\boldsymbol{v\nu})/\sqrt{1\!-\!{\boldsymbol v}^2} \!=\!const$, i.e., of surfaces perpendicular (see~Fig.\,1) to the congruence of geodesics parallel to $\boldsymbol\nu$  and possessing Euclidean inner geometry.\\
\begin{figure}[hbt]
\vspace*{-8mm}
\begin{center}
\epsfig{figure=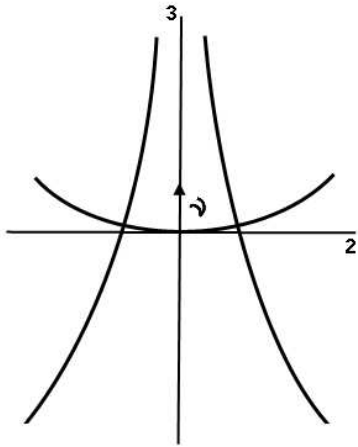,scale=0.62}
\vspace*{-2mm}
\caption{Horosphere 2D image in the Lobachevski space. The horosphere belongs to the family $(1\!- \!\boldsymbol{v\nu})/\sqrt{1\!-\!{\boldsymbol v}^2} \!=\!const.$}
\end{center}
\end{figure}

\newpage
Now, along with initial Finsler space-time (1), let us consider the Finsler space-time (2). Since its metric
can be obtained from (1) by replacing $\boldsymbol\nu\rightarrow -\boldsymbol\nu$, we shall treat (2)
as axially symmetric  Finsler space-time with the opposite direction of $\boldsymbol\nu .$

For easier comparison of corresponding equations peculiar to the spaces (1) and (2), we represent, for example, infinitesimal transformations of 3-parameter groups of relati\-vis\-tic symmetry of these spaces in the following form
\begin{eqnarray}\label{12}
dx_0^{_{(2)}^{(1)}}\!&=&\!\!(\,\mp\, r(\boldsymbol{\nu}\boldsymbol{n})x_0 - \boldsymbol{n} \boldsymbol{x}\,)\, d \alpha\,, \nonumber\\
{d\boldsymbol{x}}^{_{(2)}^{(1)}}\!\!&=&\!\!(\,\mp\, r(\boldsymbol{\nu}\boldsymbol{n})\boldsymbol{x} - \boldsymbol{n} x_0 \mp [\boldsymbol{x}[\boldsymbol{\nu}\boldsymbol{n}]]\,)\,d\alpha\,,
\end{eqnarray}
where the unit vector $\boldsymbol n$ and $\alpha $ are the group parameters.
Here and below, two-level index ${_{(2)}^{(1)}}$ in the left side of each equation
indicates that the equation relates to space (1) and space (2).
In accordance with the architecture of this index, the upper signs in the right side
of each equation correspond to the case of space (1), whereas the  lower signs
correspond to the case of space (2).\\
As for the infinitesimal transformations of the above-mentioned 1-parameter group
of rotations and 4-parameter group of space-time translations, they have the form
\begin{equation}\label{13}
{d\boldsymbol{x}}^{_{(2)}^{(1)}}=\pm\, [\boldsymbol x\boldsymbol\nu ]\,d\omega\,; \qquad (dx^i)^{_{(2)}^{(1)}}=da^i\,, \quad i=0,\,1,\,2,\,3\,.
\end{equation}
If, as before, the spatial axes are chosen so that $\boldsymbol\nu =(0,\,0,\,1)$\,,
and three successive directions (along the first-, the second- and the third axis) are chosen for $\boldsymbol n$, then the generators and the corresponding Lie algebras of the 8-parameter isometry groups of Finsler spaces (1) and (2) appear as

\vspace*{0.7cm}
\begin{equation}
\left.
\begin{array}{rcl}
X_1^{_{(2)}^{(1)}}&=&-\,(x^1p_0+x^0p_1)\mp (x^1p_3-x^3p_1)\,,\vspace*{0.3cm}\\
X_2^{_{(2)}^{(1)}}&=&-\,(x^2p_0+x^0p_2)\pm (x^3p_2-x^2p_3)\,,\vspace*{0.3cm}\\
X_3^{_{(2)}^{(1)}}&=& \mp\, r(x^0p_0+\boldsymbol x\boldsymbol p)-(x^3p_0+x^0p_3)\,,\vspace*{0.3cm}\\
R_3^{_{(2)}^{(1)}}&=& \pm\, (x^2p_1-x^1p_2)\,; \qquad \qquad \qquad \qquad\quad p_i=\partial /\partial
x^i\,.\\
\end{array}
\right.
\end{equation}

\vspace*{1.4cm}
\begin{equation}\label{15}
\left.
\begin{array}{llll}
  [X_1^{_{(2)}^{(1)}}X_2^{_{(2)}^{(1)}}]=0\,, & \quad\qquad\, [R_3^{_{(2)}^{(1)}}X_3^{_{(2)}^{(1)}}]=0\,, && \vspace*{0.3cm}\\
\left.[X_3^{_{(2)}^{(1)}}X_1^{_{(2)}^{(1)}}]\right.=\pm\, X_1^{_{(2)}^{(1)}} \,, & \quad\qquad\, [R_3^{_{(2)}^{(1)}}X_1^{_{(2)}^{(1)}}]=\pm\, X_2^{_{(2)}^{(1)}} \,, && \vspace*{0.3cm}\\
  \left.[X_3^{_{(2)}^{(1)}}X_2^{_{(2)}^{(1)}}]\right.=\pm\, X_2^{_{(2)}^{(1)}}\,, & \quad\qquad\, [R_3^{_{(2)}^{(1)}}X_2^{_{(2)}^{(1)}}]=\mp\, X_1^{_{(2)}^{(1)}}\,; & & \\
\end{array}
\right.
\end{equation}

\vspace*{-0.5cm}
\begin{equation}
\left.
\begin{array}{llll}
  \left.[p_i p_j]\right.=0\,; &&& \vspace*{0.2cm}\\
  \left.[X_1^{_{(2)}^{(1)}}p_0]\right.=p_1\,,&\ \qquad [X_2^{_{(2)}^{(1)}}p_0]=p_2\,, && \vspace*{0.2cm}\\
  \left.[X_1^{_{(2)}^{(1)}}p_1]\right.=p_0 \pm p_3\,,&\ \qquad [X_2^{_{(2)}^{(1)}}p_1]=0\,, && \vspace*{0.2cm}\\
  \left.[X_1^{_{(2)}^{(1)}}p_2]\right.=0\,, &\ \qquad [X_2^{_{(2)}^{(1)}}p_2]=p_0 \pm p_3\,, && \vspace*{0.2cm}\\
  \left.[X_1^{_{(2)}^{(1)}}p_3]\right.= \mp\, p_1\,, &\ \qquad [X_2^{_{(2)}^{(1)}}p_3]= \mp\, p_2\,, && \vspace*{0.2cm}\\
  \left.[X_3^{_{(2)}^{(1)}}p_0]\right.= \pm\, rp_0+p_3\,, &\ \qquad [R_3^{_{(2)}^{(1)}}p_0]=0\,, && \vspace*{0.2cm}\\
  \left.[X_3^{_{(2)}^{(1)}}p_1]\right.= \pm\, rp_1\,, &\ \qquad [R_3^{_{(2)}^{(1)}}p_1]= \pm\, p_2\,, && \vspace*{0.2cm}\\
  \left.[X_3^{_{(2)}^{(1)}}p_2]\right.= \pm\, rp_2\,, &\ \qquad [R_3^{_{(2)}^{(1)}}p_2]= \mp\, p_1\,, && \vspace*{0.2cm}\\
  \left.[X_3^{_{(2)}^{(1)}}p_3]\right.= \pm\, rp_3+p_0\,, &\ \qquad [R_3^{_{(2)}^{(1)}}p_3]=0\,. && \\
\end{array}
\right.
\end{equation}

\vspace*{0.5cm}
Now compare the 3-parameter noncompact homogeneous group of relativistic symmetry
of space (1) (\,the generators $X_1^{(1)}, X_2^{(1)}, X_3^{(1)}$\,) with
the corresponding group of space (2) (\,the generators
$X_1^{(2)}, X_2^{(2)}, X_3^{(2)}$\,). According to their Lie algebras (\,see (15)\,),
these groups are isomorphic 
to the corresponding 3-parameter subgroups (\,with generators $X_1^{(1)}, X_2^{(1)}, X_3^{(1)}\!\!\mid_{r=0}$  and $X_1^{(2)}, X_2^{(2)}, X_3^{(2)}\!\!\mid_{r=0},$ respectively\,)
of the homogeneous Lorentz group. Similarly to the case of space (1), the Abelian 2-parameter subgroup (\,with the generators $X_1^{(2)}, X_2^{(2)}$\,)
induces a 2-parameter subgroup of such motions of the Lobachevski space which
leave invariant a family of the horospheres $(1\!+ \!\boldsymbol{v\nu})/\sqrt{1\!-\!{\boldsymbol v}^2} \!=\!const$, i.e. of surfaces perpendicular (see~Fig.\,2)  to the congruence of geodesics parallel to
$-\boldsymbol\nu$ and possessing Euclidean inner geometry.\\
\begin{figure}[hbt]
\vspace*{-5mm}
\begin{center}
\epsfig{figure=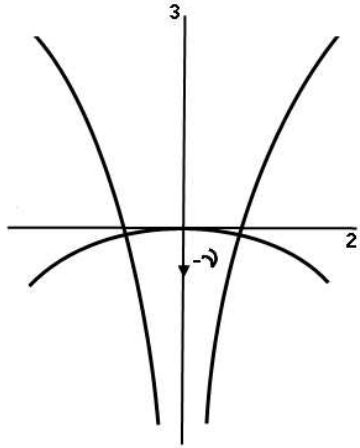,scale=0.62}
\caption{Horosphere 2D image in the Lobachevski space. The horosphere belongs to the family $(1\!+ \!\boldsymbol{v\nu})/\sqrt{1\!-\!{\boldsymbol v}^2} \!=\!const.$}
\end{center}
\end{figure}

\newpage
\noindent
{\large{\bf 3\,. Two 3-parametric noncompact subgroups of the homogeneous \phantom{Lor}Lorentz group as rudiments of the 3-parametric groups of \phantom{rel\!a}relativistic symmetry of the axially symmetric Finsler spaces \phantom{rel\!a}with mutually opposite preferred directions}}

\bigskip
\bigskip

\noindent
If $\,r=0\,,$ the metrics of axially symmetric Finsler spaces (1) and (2), i.e.
\begin{equation}\label{17}
\left (ds^2\right )^{_{(2)}^{(1)}}=\left[\frac{(dx_0\mp\boldsymbol\nu d\boldsymbol
x)^2}{dx_0^2-d\boldsymbol x^{2}}\right]^r (dx_0^2-d\boldsymbol
x^{2})\,,
\end{equation}
reduce to the Minkowski one $ds^2=dx_0^2-d\boldsymbol x^{2}\,.$ However transformations of  relativisic symmetry of these spaces, i.e. transformations
\begin{equation}\label{18}
\left (x'^{\,i}\right )^{_{(2)}^{(1)}}=D(\boldsymbol v,\pm\boldsymbol\nu )R^i_j(\boldsymbol v,\pm\boldsymbol\nu )L^j_k(\boldsymbol v) x^k\,,
\end{equation}
in which
\begin{equation}\nonumber
D(\boldsymbol v,\pm\boldsymbol\nu )=\left(\frac{1\mp\boldsymbol
v\boldsymbol\nu /c} {\sqrt{1-\boldsymbol v^{2}/c^2}} \right)^rI\,\,,
\end{equation}
do not reduce to the ordinary Lorentz boosts
\begin{equation}\label{19}
x'^{\,i}=L^i_k(\boldsymbol v) x^k\,.
\end{equation}
Incidentally, these boosts can be represented in the following explicit form
\begin{eqnarray}\nonumber
x'_0=\frac{x_0-(\boldsymbol v\boldsymbol x)}{\sqrt{1-{\boldsymbol v}^2}}\,,\\\label{20}
{\boldsymbol x}'=\boldsymbol x-\frac{\boldsymbol v}{\sqrt{1-{\boldsymbol v}^2}}\left[x_0-\left(1-\sqrt{1-{\boldsymbol v}^2}\right)\left(\boldsymbol v\boldsymbol x\right)/{\boldsymbol v}^2\right]\,.
\end{eqnarray}

At $\,r=0\,,$ i.e. in the case of Minkowski space where all directions in 3D space are equivalent,
 $\,\boldsymbol\nu\,$ has no physical meaning. In this case, each of the two  rudimentary transformations
\begin{equation}\label{21}
x'^{\,i}=R^i_j(\boldsymbol
v,\pm\boldsymbol\nu )L^j_k(\boldsymbol v) x^k
\end{equation}
differs from  the Lorentz boost (19) by the corresponding additional rotation
\begin{equation}\label{22}
x'^{\,i}=R^i_k(\boldsymbol v,\pm\boldsymbol\nu ) x^k
\end{equation}
of the spatial axes. This additional rotation is adjusted in such a way that if a ray of light has the direction $\,\boldsymbol\nu\,$ or $\,-\boldsymbol\nu\,$ in one frame, then it  will have respectively the same direction in all the frames.

In order to find an explicit form of the additional rotation (22) we should use the following general formula
\begin{equation}\label{23}
{\boldsymbol x}'=\boldsymbol x+[\boldsymbol N[\boldsymbol N\boldsymbol x]](1-\cos\varphi )-[\boldsymbol N\boldsymbol x]\sin\varphi\,.
\end{equation}
This formula determines the transformed components ${\boldsymbol x}'$ of radius vector $\boldsymbol x$ after rotation of the spatial axes around arbitrary unit vector $\boldsymbol N$ through an angle $\varphi$.\\

\newpage
\noindent
In our case the respective $\boldsymbol N$ and $\varphi$ can be obtained by means of solving hyperbolic triangles in the Lobachevski 3-velocity space (\,see Fig.\,3).
\begin{figure}[hbt]
\hspace*{-1.5cm}
\begin{center}
\epsfig{figure=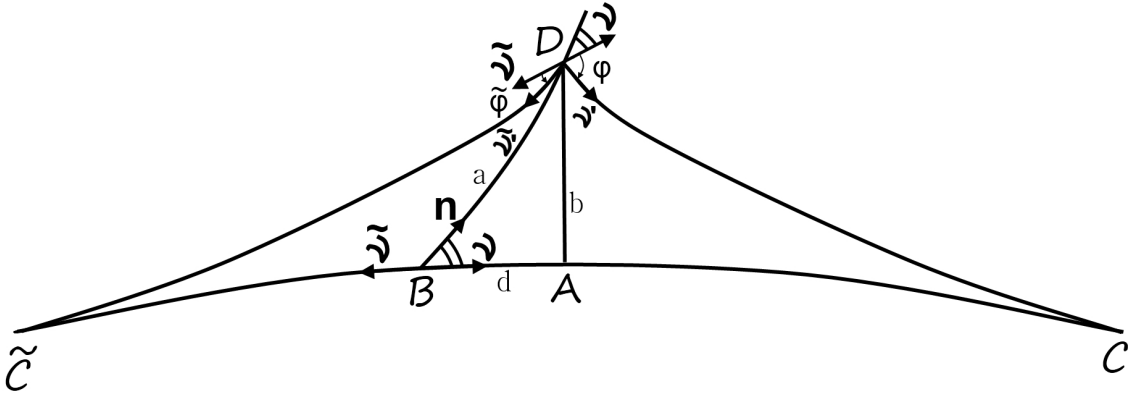,scale=0.31}
\end{center}
\vspace*{-0.5cm}
\caption{Hyperbolic triangles in the relativistic 3-velocity space.}
\end{figure}

\vspace*{0.9cm}
In Fig.\,3, the point $\,\cal B\,$ depicts the initial reference frame, $\cal D$ the reference frame moving at velocity $\boldsymbol v$ (\,the unit vector $\boldsymbol n$ indicates the direction of $\boldsymbol v$, i.e. $\boldsymbol n={\boldsymbol v}/v$\,). In the reference frame $\cal B$ the ray of light has the direction $\,\boldsymbol\nu\,$ or $\,\boldsymbol{\tilde\nu}=-\boldsymbol\nu\,$, and in the reference frame $\cal D$, the direction $\,\boldsymbol{{\nu}'}\,$ or $\,\boldsymbol{{\tilde\nu}'}\,$, respectively. The $\cal{DC}\tilde{\cal C}$ angle is zero (\,the straight lines $\,\cal{DC}\,$ and $\,\tilde{\cal C}\cal C\,$ are parallel\,). The $\cal D\tilde{\cal C}\cal C$ angle is zero (\,the straight lines $\,\cal D\tilde{\cal C}\,$ and $\,\cal C\tilde{\cal C}\,$ are also parallel\,). In addition, $\angle{\cal{DAC}}=\angle{\cal{DA}\tilde{\cal C}}=\pi /2 $ and $\angle{\cal{ADC}}=\angle{\cal{AD}\tilde{\cal C}}=\Pi (b) $. Here $\,\Pi (b)\,$ is the Lobachevski angle for parallelism. Its dependence on the distance $\,b\,$ between two points $\,\cal D\,$ and $\,\cal A\,$ is determined by the formula $\,\Pi (b)=2\arctan{e^{-b}}\,.$

In order to find the angle $\varphi$ of relativistic aberration of $\,\boldsymbol\nu\,$, the vector $\,\boldsymbol\nu\,$ should be carried in parallel from $\,\cal B\,$ to $\cal D$ along the straight line $\,\cal{BD}\,$ and then one should make use of the formulae of hyperbolic
geometry, taking into account that $\,\tanh a=v/c\,$. As a result,
we arrive at (9). From the same Fig.\,3 one can see that the rotation is performed around the vector $\,[\boldsymbol v\boldsymbol\nu ]\,$. Now, with the help of (23), we are able to write down the transformation which corresponds to the additional rotation $\,x'^{\,i}=R^i_k(\boldsymbol v, \boldsymbol\nu ) x^k\,$ of the spatial axes. It has the form
\begin{eqnarray}\nonumber
{\boldsymbol x}'=\boldsymbol x+\frac{(\sqrt{1-{\boldsymbol v}^2}-1)(\boldsymbol v\boldsymbol x)+{\boldsymbol v}^2(\boldsymbol\nu\boldsymbol x)}{{\boldsymbol v}^2(1-\boldsymbol v\boldsymbol\nu )}\,\boldsymbol v+\phantom{+\sqrt{1-{\boldsymbol v}^2}+(1)}\\\label{24}
+\,\,\frac{(1-\sqrt{1-{\boldsymbol v}^2})[2(\boldsymbol v\boldsymbol\nu )(\boldsymbol v\boldsymbol x)-{\boldsymbol v}^2(\boldsymbol\nu\boldsymbol x)]-{\boldsymbol v}^2(\boldsymbol v\boldsymbol x)}{{\boldsymbol v}^2(1-\boldsymbol v\boldsymbol\nu )}\,\boldsymbol\nu \,.
\end{eqnarray}
Note that in (24) we put $\,c=1\,$.

Similarly, in order to find the angle $\tilde{\varphi}$ of relativistic aberration of $\,\boldsymbol{\tilde\nu}\,$, the vector $\,\boldsymbol{\tilde\nu}\,$ should be carried in parallel from $\,\cal B\,$ to $\cal D$ along the straight line $\,\cal{BD}\,$. As a result, we get
\begin{equation}\label{25}
\tilde{\varphi}=\arccos\left\{ 1-\frac{(1-\sqrt{1-\boldsymbol v^{2}/c^2})
[\boldsymbol\nu\boldsymbol v]^2}{ (1+\boldsymbol v\boldsymbol\nu
/c)\boldsymbol v^{2}}\right\}
\end{equation}
From Fig.\,3 one can see that such a rotation is performed around the vector $\,[\boldsymbol\nu\boldsymbol v]\,$.
Now, with the help of (23) we are able to write down the transformation which corresponds to another additional rotation $\,x'^{\,i}=R^i_k(\boldsymbol v, \boldsymbol{\tilde\nu} ) x^k=R^i_k(\boldsymbol v, -\boldsymbol\nu ) x^k\,$ of the spatial axes. It has the form
\begin{eqnarray}\nonumber
{\boldsymbol x}'=\boldsymbol x+\frac{(\sqrt{1-{\boldsymbol v}^2}-1)(\boldsymbol v\boldsymbol x)-{\boldsymbol v}^2(\boldsymbol\nu\boldsymbol x)}{{\boldsymbol v}^2(1+\boldsymbol v\boldsymbol\nu )}\,\boldsymbol v+\phantom{+\sqrt{1-{\boldsymbol v}^2}+(1)}\\\label{26}
+\,\,\frac{(1-\sqrt{1-{\boldsymbol v}^2})[2(\boldsymbol v\boldsymbol\nu )(\boldsymbol v\boldsymbol x)-{\boldsymbol v}^2(\boldsymbol\nu\boldsymbol x)]+{\boldsymbol v}^2(\boldsymbol v\boldsymbol x)}{{\boldsymbol v}^2(1+\boldsymbol v\boldsymbol\nu )}\,\boldsymbol\nu \,.
\end{eqnarray}
Here, as before, we put $\,c=1\,$.
\vskip 5mm
\noindent
{\large{\bf 4\,. Conclusion}}

\bigskip

\noindent
Having studied the axially symmetric Finsler spaces with mutually opposite preferred directions and their
isometry groups, we gave particular attention to  the limiting case $\,r=0\,$. As it turned out, if $\,r=0\,,$ the respective Finsler metrics $\,ds^2=\left[(dx_0\mp\boldsymbol\nu d\boldsymbol
x)^2/(dx_0^2-d\boldsymbol x^{2})\right]^r (dx_0^2-d\boldsymbol x^{2})\,$ reduce to the Minkowski one $\,ds^2=dx_0^2-d\boldsymbol x^{2}\,.$ However, transformations of  relativistic symmetry of the above-mentioned Finsler spaces
do not reduce to the ordinary Lorentz boosts. At $\,r=0\,,$ i.e. in the case of Minkowski space where all directions in 3D space are equivalent, $\,\boldsymbol\nu\,$ has no physical meaning. In this case, each of the transformations of the pair of rudimentary transformations $\,x'^{\,i}=R^i_j(\boldsymbol v,\pm\boldsymbol\nu )L^j_k(\boldsymbol v) x^k\,$ differs from  the Lorentz boost $\,x'^{\,i}=L^i_k(\boldsymbol v) x^k\,$ by the corresponding additional rotation $\,x'^{\,i}=R^i_k(\boldsymbol v,\pm\boldsymbol\nu ) x^k\,$ of the spatial axes. This additional rotation is adjusted in such a way that if a ray of light has the direction $\,\boldsymbol\nu\,$ or $\,-\boldsymbol\nu\,$ in one frame, then it  will have respectively the same direction in all the frames. Thus, at $\,r=0\,,$ the two sets of rudimentary transformations represent two alternatives to the Lorentz boosts, however, in contrast to the boosts, they constitute two different but isomorphic 3-parameter noncompact groups (\,see the two horospheres in Fig.\,4 illustrating this fact\,).\\
Physically, such noncompact transformations are realized as follows. First choose as $\,\boldsymbol\nu\,$ a direction towards a preselected star (\,or opposite direction\,) and then perform an arbitrary Lorentz boost by complementing it with such a turn of the spatial axes that in a new reference frame the direction towards the star (\,or opposite direction, respectively\,) remains unchanged. These two sets consisting of the described transformations form two different 3-parameter noncompact groups.
\begin{figure}[hbt]
\vspace*{-0.4cm}
\begin{center}
\epsfig{figure=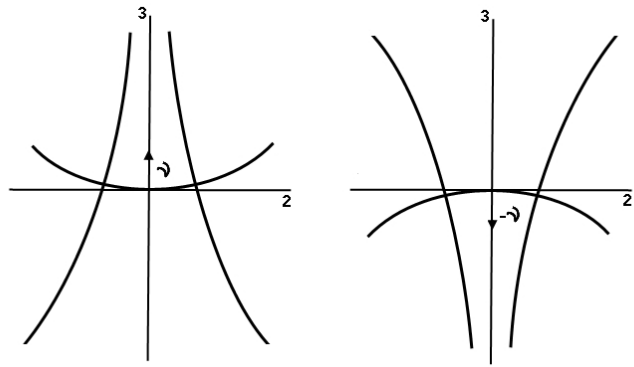,scale=0.5}
\end{center}
\vspace*{-1.0cm}
\caption{Two different horospheres in Lobachevski space as examples of two different surfaces of transitivity arising from the two rudimentary groups.}
\end{figure}

\end{document}